\input ./style/arxiv-general.cfg
\documentclass[aoas,MSNbibl,nameyear,noautosecdot,rotating,dvips]{arximspdf}
\makeatletter
   \@ifpackageloaded{graphicx}{}{\usepackage{graphicx}}
\makeatother
\usepackage{url,breakurl}

\doi{10.1214/15-AOAS858}
\volume{9}
\issue{4}
\pubyear{2015}
\firstpage{1950}
\lastpage{1972}
\docsubty{FLA}

\makeatletter
\newcommand{\rrVert}{\Vert}
\newcommand{\llVert}{\Vert}
\makeatother

\begin{document}
\begin{frontmatter}

\title{Analysis of multiview legislative networks with structured
matrix factorization: Does Twitter influence translate to the real world?}
\runtitle{Analyzing multiview networks with matrix factorization}

\begin{aug}
\author[A]{\fnms{Shawn}~\snm{Mankad}\corref{}\ead[label=e1]{smankad@cornell.edu}}
\and
\author[B]{\fnms{George}~\snm{Michailidis}\ead[label=e2]{gmichail@umich.edu}}
\runauthor{S. Mankad and G. Michailidis}
\affiliation{Cornell University and University of Michigan}
\address[A]{Operations, Technology and Information Management\\
Cornell University\\
Ithaca, New York 14850\\
USA\\
\printead{e1}}
\address[B]{Department of Statistics\\
University of Michigan\\
Ann Arbor, Michigan 48109\\
USA\\
\printead{e2}}
\end{aug}

%
\received{\smonth{10} \syear{2014}}
%
\revised{\smonth{7} \syear{2015}}

%
\begin{abstract}
The rise of social media platforms has fundamentally altered the
public discourse by providing easy to use and ubiquitous forums for
the exchange of ideas and opinions. Elected officials often use such
platforms for communication with the broader public to disseminate
information and engage with their constituencies and other public
officials. In this work, we investigate whether Twitter conversations
between legislators reveal their real-world position and influence by
analyzing multiple Twitter networks that feature different types of
link relations between the Members of Parliament (MPs) in the United
Kingdom and an identical data set for politicians within Ireland. We
develop and apply a matrix factorization technique that allows the
analyst to emphasize nodes with contextual local network structures by
specifying network statistics that guide the factorization solution.
Leveraging only link relation data, we find that important politicians
in Twitter networks are associated with real-world leadership
positions, and that rankings from the proposed method are correlated
with the number of future media headlines.
\end{abstract}

%
\end{frontmatter}

\section{Introduction\texorpdfstring{.}{}}

There is a growing literature that attempts to understand and exploit
social networking platforms for resource optimization and marketing,
as it is a major interest for private enterprises and political
campaigns attempting to propagate particular opinions or products
[\citeauthor{TwitterNYT3} (\citeyear{TwitterNYT3,TwitterNYT2,TwitterNYT})]. An important problem is
the identification of influential individuals that facilitate
communication over the network. In this paper, we develop a modeling
approach that captures influence from multiple networks that feature
different link relations between the same set of nodes (e.g., Twitter
accounts). Such multiview data are increasingly common due to the
complex structure of many networking platforms. Specifically, we
analyze three different types of networks that are commonly derived
from Twitter data, each composed of either weighted or binary links.

Twitter is a popular platform with over 270 million active accounts
each month as of September 2014 [\citet{TwitterStatistics}]. Twitter
allows accounts to post short messages of 140 characters or less,
commonly referred to as ``tweets,'' that can be read by any visitor. A
tweet that is a copy of another account's tweet is called a
``retweet.'' Within a tweet, an account can mention another account by
referring to their account name with the $@$ symbol as a
prefix. Accounts also declare the other accounts they are interested
in ``following,'' which means the follower receives notification
whenever a new tweet is posted by the followed account. These three
directed actions define political Twitter networks that we analyze in
this work.

The first network is a retweet network, where links are
directed and weighted to denote the log-number of retweets from one
account to another over an interval of time.
The second network is also composed of directed
and weighted links that denote the log-number of mentions one account
gives another. The third network is constructed with directed binary
links that denote the follower and followed relationships between
accounts.

\begin{figure}
\begin{tabular}{@{}ccc@{}}

\includegraphics{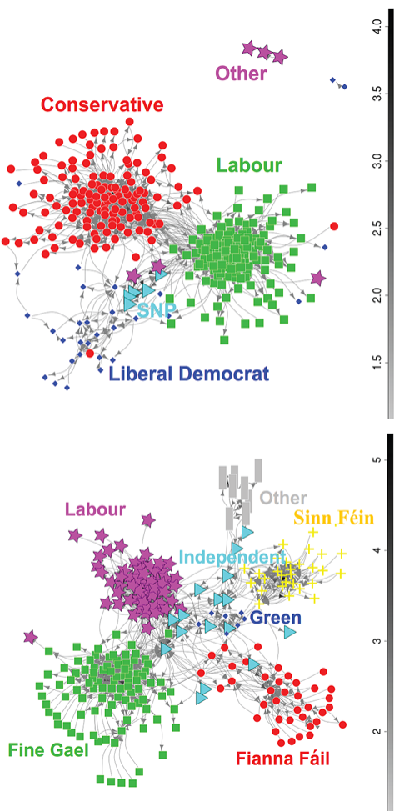}
 & \includegraphics{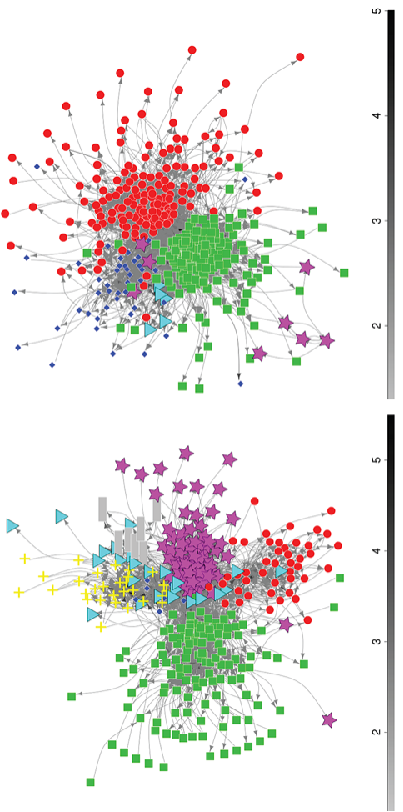} &\includegraphics{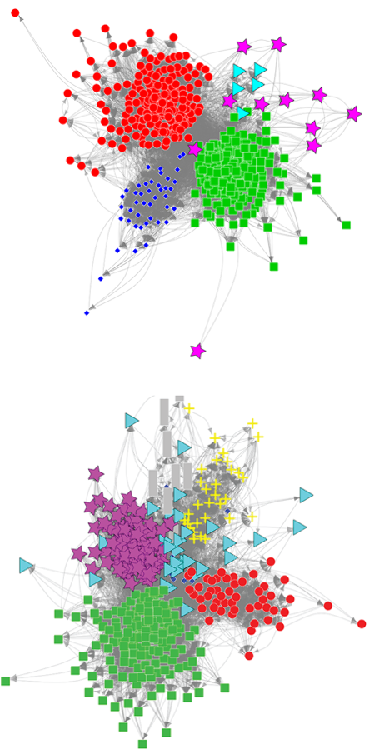}\\
\footnotesize{(a) Retweet network} & \footnotesize{(b) Mentions
network} & \footnotesize{(c) Follows network}
\end{tabular}
\caption{The top panel shows networks of UK Members of Parliament and
the bottom panel
shows networks of Irish politicians and political organizations. Node
color and vertex shapes denote party affiliation. The average degree
for the UK Retweet, Mentions and Follows network is 9.13,
25.51 and 65.25, respectively. The average degree for the Irish
Retweet, Mentions and Follows network shown in the bottom row is
5.81, 15.28 and 48.44, respectively.}
\label{fig:UKTwitter}
\end{figure}

These three networks, each featuring 416 Members of Parliament (MPs)
in the United Kingdom, are drawn in the top panel of
Figure~\ref{fig:UKTwitter}, where accounts are registered to 172
Conservative MPs, 185 Labour, 43 Liberal Democrats, 5 MPs representing
the Scottish National Party (SNP), and 11 MPs belonging to other
parties. There are 650 MPs forming the House of Commons, the lower
house in the bicameral legislative body for the United Kingdom. Each
MP is democratically elected to represent constituencies for five year
terms, though often elections are held more frequently when Parliament
is dissolved.

The second set of political Twitter networks that we analyze are drawn
in the bottom panel of Figure~\ref{fig:UKTwitter}. Each network is
composed of 348 nodes that represent the accounts of Irish politicians
and political organizations at all levels of government, including the
President of the Republic of Ireland, members of the local and
national government, and elected representatives for the European
Union.

The raw data for both data sets, collected and processed by
\citet{2013arXiv1301.5809G}, consists of approximately 500,000 tweets
and 40,000 follower links from late 2012. An empirical pattern
observed in these data and also in previous studies [\citet{hpTwitter}]
is that the follower network is very dense in contrast to the retweet
and mentions networks. Almost all politicians interact via retweeting
or mentioning with a smaller number of other
accounts, relative to their follower declarations. Moreover, users
with many followers post updates less often than those with fewer
followers [\citet{hpTwitter}]. Such empirical findings suggest
that not
all links are created equally, and usually the follower network is
discarded because it does not accurately capture patterns of conversation.
However, each network, including the follower
network, contains meaningful information, especially since we only
consider the population of politicians in a specific legislative body
instead of a broad set of users or even the entire Twitter userbase.

%

Previous research has found that Twitter and other social networking
platforms help facilitate communication between politicians,
government agencies and the broader public. \citet{TwitterCongress}
find by text mining tweets that members of the United States Congress
employ Twitter for primarily two purposes: information dissemination
and self promotion. \citet{predictionElectionsTwitter} find that the
number of tweets from the general public mentioning a political party
or politician is a valid indicator of political sentiment and a good
predictor of federal election results in Germany. More recently,
similar results have been found for federal elections in Australia and
the U.S. House of Representatives
[\citet
{predictionElectionsTwitterWISE,predictionElectionsTwitter2014}]. In
contrast to these previous
works, we rely only on the link relations, so-called ``meta-data,''
among politicians to measure influence and identify conversation flows with
network analysis. Approaches that utilize content analysis can face
significant challenges associated with text and image analysis
(accounts can post a photo within a tweet), such as language
differences, tone and sentiment characterization, and so on.

There has been extensive work on ranking nodes on a network by their importance
primarily motivated by search on the World Wide Web. We find our
proposed method compares favorably for ranking politicians against two
seminal works called PageRank [\citet{page1999pagerank}] and HITS
[Hyperlink-Induced Topic Search; \citet{kleinberg-authority}].
The idea
behind PageRank is to use as a measure of importance an estimate of
the probability of reaching a given node by randomly following
edges. HITS utilizes the so-called authority and hub scores, which are
computed by the leading eigenvector of $A^{T}A$ and $AA^{T}$,
respectively, where $A$ is an adjacency matrix.

Our main goal of identifying influential politicians is also closely
related to role identification, which aims to assign roles based on
local connectivity patterns. Typically, role analysis methods rely on
analyzing ego networks (the union of a node and its neighbors),
network statistics or graph-coloring techniques
[\citet{JCGSRoleAnalysis}]. Also note that while there have been many
recent advances in community detection, including the stochastic block
model, latent position cluster models and others [see
\citet{Fienberg-networkOverview-jcgs,SAMSAM11146} for survey
articles], the task in this article is different from typical
community detection, which aims to extract groups of nodes that
feature relatively dense within-group connectivity and sparser
between-group connectivity. That said, community detection could help
guide a
search for influential politicians. For instance, an analyst may
examine each network separately by first discovering communities, if
unknown, then searching for interesting network statistic profiles
within each group. There are in principle many ways to combine
community detection with network statistics for the identification of
influential nodes, (e.g., politicians), but it remains unclear which
is the preferred method. In this paper, we integrate both steps
together to address this issue. The proposed factorization model is
also able to emphasize nodes with interesting path-related properties
by incorporating node-level statistics that capture these nonlinear
relationships, thus leading to more interpretable measures of
influence and substructure.

The main idea is to guide the mapping of the multiview networks to
lower-dimensional spaces using structured matrix factorization.
Nonnegativity constraints are also imposed on the lower-dimensional
spaces to improve data representation and structural discovery. Such
constraints have been popularized with the nonnegative matrix
factorization (NMF) and Semi-NMF, where one or all matrix factors are
composed of only nonnegative entries and have been shown to be
advantageous for data representation [\citet{NNMF1,ConvexNMF}]. As
validation, we find that important politicians identified using our
modeling approach are associated with real-world leadership positions,
and that rankings from the proposed method are significantly
correlated with future media headlines. The consistent findings
between both data sets suggest the model can be a relatively
straightforward technique for identifying influential individuals with
political Twitter networks from other countries that feature
different government structures, and that it can complement the
potentially more involved content analysis for related tasks.

The next section introduces the matrix factorization model, followed
by estimation details in Section~\ref{sec:algo}. Section~\ref
{sec:nmfResults} summarizes and compares results of the proposed
model against alternative methodologies with UK MPs and Irish
politicians. This article closes with a brief discussion in
Section~\ref
{sec:conclusion}.

\section{Structured semi-NMF for influence discovery\texorpdfstring
{.}{}} \label{sec:model}
The use of low-rank approximations to network related matrices follows
a long line of previous work. In classical spectral layout, the
coordinates of each node are given by the Singular Value Decomposition
(SVD) of the Laplacian matrix
[\citet{GraphDrawing-Koren,Dynamic-Spectral-Drawing}]. Recently, there
has been extensive interest in spectral clustering
[\citet{2012arXiv1204.2296R,1227.62042}], which
discovers community structure in the eigenvectors of the Laplacian matrix.

Low-rank approximations satisfying different constraints other than
orthonormality are also popular. For instance, NMF has been proposed
for overlapping community detection on static
[\citet{bayesNMF-networks,Ding2}] and dynamic [\citet
{facetnet}] networks.
When overlaps among communities exist, an advantage of NMF over
spectral clustering is that NMF can still find basis vectors for each
community, while orthogonality of SVD makes it unlikely that the
singular vectors will correspond to each of the communities
[\citet{xu2003}]. The basic framework for NMF in network analysis
is $A
\approx UV^{T}$, where $A$ is an adjacency matrix and $U,V \in
\mathbb{R}_{\ge0}^{n\times K}$. Written in element form,
\[
A_{ij} \approx U_{i1} V_{j1} + \cdots+
U_{iK}V_{jK},
\]
one can easily see that each edge of the given network is approximated
with a nonnegative sum. Consequently, each term in the sum,
$U_{ik}V_{jk}$, represents the contribution of the $k$th
latent structure (often capturing community structure especially
when decomposing sparse adjacency matrices [\citet{PhysRevE.88.042812}])
to the edge from $i$ to $j$. Edge decompositions can be aggregated by
node or one can use the rows of $V$ to directly determine node community
membership. The factors are found by minimizing
\[
\min_{U\ge0,V\ge0}\bigl\llVert A - UV^{T}\bigr\rrVert
_{F}^{2},
\]
where $\llVert\cdot\rrVert_{F}$ denotes the Frobenius norm. The
optimization can
be performed using gradient-descent algorithms for penalized
optimization. Given that the proposed model in this article utilizes
nonnegativity, we follow a similar algorithmic approach to the NMF
literature.

Enforcing nonnegativity on a single matrix factor was first proposed
in \citet{ConvexNMF} with the so-called Semi-NMF to improve
interpretability of the resultant factorizations with data of mixed
signs. We utilize the flexibility of Semi-NMF and extend it to the
network setting with a structured approach that incorporates graph
geometry into the factorization through user-specified matrices. In
particular, we aim to utilize the many node-level statistics that have
been proposed in the network literature to guide the factorization
solution. Next we introduce the model for singleview networks,
then extend to multiview networks, followed by estimation procedures
in the next section.

\subsection{Singleview networks\texorpdfstring{.}{}}
Let $A$ denote the adjacency matrix from a single, given network with
$n$ nodes. We start with the following graph Structured Semi-NMF model
of \citet{6714064}:
%
\begin{equation}
\label{eqn:objective:single} \min_{\Lambda, \Theta\ge0} \bigl
\llVert
A - S\Lambda
\Theta^{T}\bigr\rrVert_{F}^{2},
\end{equation}
where $S\in\mathbb{R}^{n\times D}, \Lambda\in\mathbb{R}^{D\times
K}$, and
$\Theta\in\mathbb{R}_{\ge0}^{n\times K}$. Note that $\Theta$ is
nonnegatively constrained, but $\Lambda$ is not, which is why the
model fits into the Semi-NMF framework. Each factor in the product
$\Lambda\Theta^{T}$ is estimated from the data and provides
coefficients for each node that represent the given adjacency matrix
in terms of $S$.

The $S$ matrix is composed of $D$ node-level statistics that are
specified by the analyst before performing the factorization to
emphasize nodes that drive influence. There is an
extensive literature in network analysis providing potential
node-level statistics [\citet{newman2010networks}].
In our analysis, the $S$ matrix is constructed using $D=4$ network statistics
and has form
\[
S_{i} = [\mbox{clustering coefficient}_{i},
\mbox{betweenness}_{i},\mbox{closeness}_{i},
\mbox{degree}_{i} ],
\]
where $i=1,\ldots,n$. The \emph{clustering
coefficient} for a given node quantifies how close its neighbors are
to forming a complete graph [\citet{newman2010networks}]. A higher
clustering coefficient
will emphasize politicians that ``create buzz.'' \emph{Betweenness}
[\citet{freeman1979centrality}]
and \emph{closeness} [\citet{newman2010networks}] rely on shortest
path statistics and
capture important links from hub nodes. \emph{Degree}, the number of
connections a
node has obtained, ensures that active politicians within communities
are emphasized in the
factorization.


If there are no node-specific values that are obvious to use for $S$,
one can start with many candidate node-level statistics and search for
subsets that fit the data well while maintaining interpretability.
This strategy will be discussed further below to also show robustness
and assess the specification of $S$ in our application. Instead of
searching over node-specific statistics, one could also be tempted to
set $S=I_{n\times n}$ to be the identity matrix. In this case, the
factorization is essentially the standard Semi-NMF factorization. Our
results show that the Semi-NMF model performs similarly to classical
importance measures, like PageRank and HITS, which should be preferred
due to their more efficient implementations.

The proposed model implies certain connectivity dynamics that can be
seen when equation~(\ref{eqn:objective:single}) is written in element
form
\begin{eqnarray*}
A_{ij} & \approx& (S\Lambda)_{i1}\Theta_{j1} +
\cdots+ (S\Lambda)_{iK}\Theta_{jK},
\\
(S\Lambda)_{ik} & = & S_{i1}\Lambda_{1k} + \cdots
+ S_{iD}\Lambda_{Dk}.
\end{eqnarray*}
For any node $i$, outgoing edges are controlled by its local
topological characteristics, as measured in $S$, and how communities
load onto the statistics in $S$, given in the columns of $\Lambda$.
When multiplied together, $S\Lambda$ form centroids in a $K$-dimensional
space that capture the outgoing node influence from each of the
communities. The receiving node $j$ in an edge is determined by the
$j$th row of $\Theta$, where larger values mean the node is more
likely to have incoming connections and, hence, greater influence.

Due to nonnegativity and the fact that $\Theta$ modulates incoming
connections, we accomplish our ultimate goal of measuring overall
influence for the $i$th node by taking its cumulative sum of
importance to each community
%
\begin{equation}
\label{eqn:importance} \mathcal{I}_{i}=\sum_{k=1}^{K}
\Theta_{ik}.
\end{equation}
As illustrated in the supplemental article [\citet
{SupplementInfo}] on a toy example,
the $S$ matrix plays a pivotal role in the factorization, and causes
$\mathcal{I}$ to be an effective importance measure even with its
relatively simple definition.

Next we propose an extension of this model to the multiview setting
found in political
Twitter networks.

\subsection{Multiview networks\texorpdfstring{.}{}}
Let $A_{m}$ denote the adjacency matrix from the corresponding Twitter
network, where $m=\{\mathrm{retweet},\mathrm{mentions},\mathrm
{follows}\}$.
We extend the singleview model with
%
\begin{equation}
\label{eqn:objective:multi} \min_{\Lambda_{m}, \Theta\ge0, V_{m}
\ge
0} \sum_{m}
\bigl\llVert A_{m} - S_{m}\Lambda_{m} (
\Theta+V_{m})^T \bigr\rrVert_{F}^2,
\end{equation}
where\vspace*{1pt} $S_{m}\in\mathbb{R}^{n\times D},
\Lambda_{m}\in\mathbb{R}^{D\times K}$, and $\Theta, V_{m}
\in\mathbb{R}_{\ge0}^{n\times K}$. $\Theta$ is common to all $m$
networks to capture general structure and makes the objective function
nonseparable, whereas $V_{m}$ reveals network-specific structure and
also implicitly weights each network according to its importance in
the factorization.

The $S_{m}$ matrices are defined similarly to the singleview case,
using node-level network statistics. We define $S_{m}$ using the same
four network
statistics for each network view. Weighted versions of the clustering
coefficient and
degree are utilized for the Retweet and Mention networks in order to
take into account the frequency of interaction between politicians,
since the
frequency should help measure the strength of a relationship
[\citet{WeightedNetworkAnalysis}]. For instance, a weighted network
statistic will distinguish between a politician that is retweeted by
the same
account hundreds of times versus retweeted once.
The model does allow for different statistics to be defined with
each network view, which may be advantageous in other contexts.

The final importance measure $\mathcal{I}$ can also be calculated
similarly using equation~(\ref{eqn:importance}). Since $\Theta$ is
common to all networks, the importance measure is a result of
integrating multiple network views in addition to structured
discovery.

\section{Algorithms\texorpdfstring{.}{}} \label{sec:algo}
The estimation algorithm we present is an iterative one that cycles
between optimizing with respect to $\Theta, V_{m}$ and $\Lambda_{m}$
with the following updates:
\begin{eqnarray}
\Theta&=& \sum_{m}A_{m}^T
S_{m} \Lambda_{m} \bigl(\Lambda_{m}^{T}S_{m}^{T}S_{m}
\Lambda_{m}\bigr)^{-1},
\nonumber
\\
V_{m} &=& A_{m}^T S_{m}
\Lambda_{m} \bigl(\Lambda_{m}^{T}S_{m}^{T}S_{m}
\Lambda_{m}\bigr)^{-1},
\nonumber
\\
\Lambda_{m} &=& \bigl(S_{m}^{T}S_{m}
\bigr)^{-1}S_{m}^{T}A_{m}(
\Theta+V_{m}) \bigl((\Theta+V_{m})^{T}(
\Theta+V_{m})\bigr)^{-1}.
\nonumber
\end{eqnarray}
The updates are based on alternating least squares (ALS) and
derived through standard arguments [\citet{ALS1}], which are
shown in the supplemental article [\citet{SupplementInfo}].

Technically, both $\Theta$ and $V_{m}$ require solving nonnegatively
constrained
least squares problems, which result in high iteration costs. So, instead
of exactly solving the constrained least squares problem, we follow a
heuristic that solves for an unconstrained solution, then sets any
entry less than a user-specified constant to that constant.
Projecting to a small constant instead of zero follows the discussion
in \citet{2008arXiv0810.4225G} and \citet{camsap2013-nmf}
to overcome
numerical instabilities that occur when too many elements are exactly
zero.

Theoretical properties are difficult to obtain due to the projection
step. Yet this approximation is computationally efficient, easy to
implement, and has been shown to achieve high quality solutions
[\citet{BerrySurvey}]. The algorithm easily scales to networks with
tens of thousands of nodes. For even larger networks on the order of
millions of nodes, low-rank factorizations should be found using
recent algorithmic advances that exploit parallel computing
architecture [\citet
{Gemulla2011LMF2020408.2020426,paralleldescent}]. For our data, we find
that the alternative least
squares algorithm is straightforward to implement and able to recover
meaningful factorizations in a timely fashion.

In the supplemental article [\citet{SupplementInfo}], we also
discuss an
alternative updating approach for $\Theta$ and $V_{m}$ that is similar
to the popular ``multiplicative updating'' for NMF. While this
approach is also very easy to implement, we find the ALS algorithm
more numerically stable in higher dimensions.

\subsection{Initialization and convergence criteria\texorpdfstring
{.}{}} \label{sec:initialization}
An advantage of the ALS algorithm is that only $\Lambda_{m}$ needs to
be initialized if the order of the updates is $\Theta, V_{m},
\Lambda_{m}$. Moreover, recall that $\Lambda_{m}$ is unconstrained,
thus bypassing the difficulties of initializing the nonnegative
factors which have received extensive focus in the NMF literature. We
find stable results by initializing $\Lambda_{m}$ with normally
distributed entries having unit mean and variance.

Another important issue is specifying the rank of the matrices
$\Theta$ and $V_{m}$. Ideally, the rank should be equal to the number
of underlying communities and can be ascertained by examining the
accuracy of the reconstruction as a function of rank. In principle,
one could also apply cross-validation procedures for matrix
factorization [\citet{owenCV}], though this may become cumbersome with
sparse or extremely large-sized networks.

\begin{figure}[t]

\includegraphics{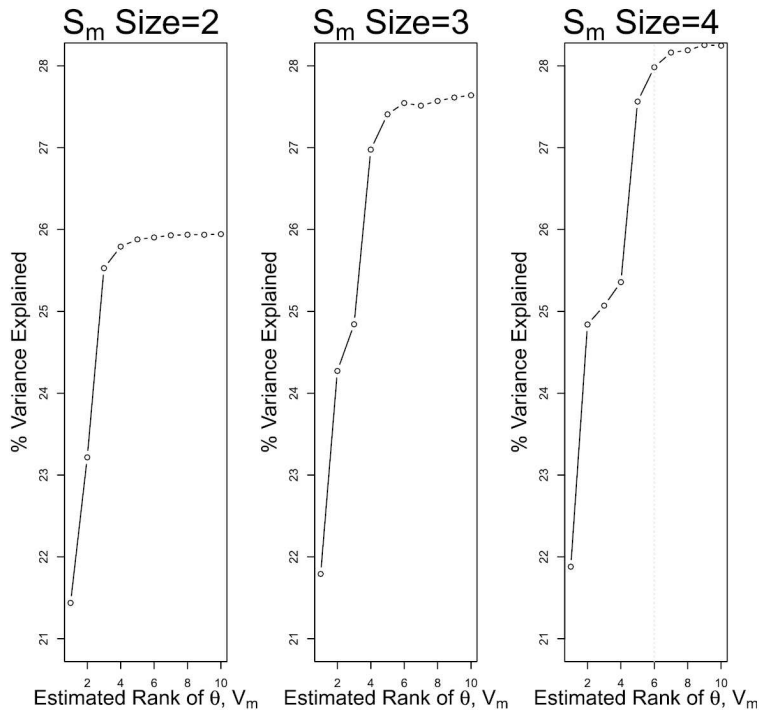}

\caption{The percentage of variance explained [$100*(1 -
\sum_{m}\llVert A_{m} - \hat{A}_{m}\rrVert^{2}_{F} / \llVert A_{m} -
\hat{\mu}_{m}\rrVert_{F}^{2})$, where $\hat{\mu}$ is a matrix filled
with the average value of $A_{m}$] for the Structured Semi-NMF with
different constructions of $S_{m}$. Plotted is the most accurate model
over thirty trials with random initializations for $\Lambda_{m}$ at
each possible specification.
We use the best rank six model with four network statistics composing
$S_{m}$ for the final analysis.}
\label{fig:UKTwitter:rank}
\end{figure}

We follow a strategy similar to using a scree plot to choose the
number of components to retain in Principal Component Analysis
[\citet{jolliffe1986principal}]. To our knowledge, this rank selection
approach has not been previously pursued in the context of NMF or
Semi-NMF. Shown in Figure~\ref{fig:UKTwitter:rank}, we find that
ranks greater than six (roughly the number of underlying political parties)
yield little marginal explanatory power. Each subfigure is constructed by
plotting the best fitting factorization over all possible network
statistic subsets of size two through four. The appropriate rank of the
matrices $\Theta$ and
$V_{m}$ is stable across the $S_{m}$ subsets, though there appears to be
significant improvement when $S_{m}$ is defined with at least three of the
network statistics. We keep all four network statistics when defining $S_{m}$
for our analysis.

%



%
\begin{figure}[t]

\includegraphics{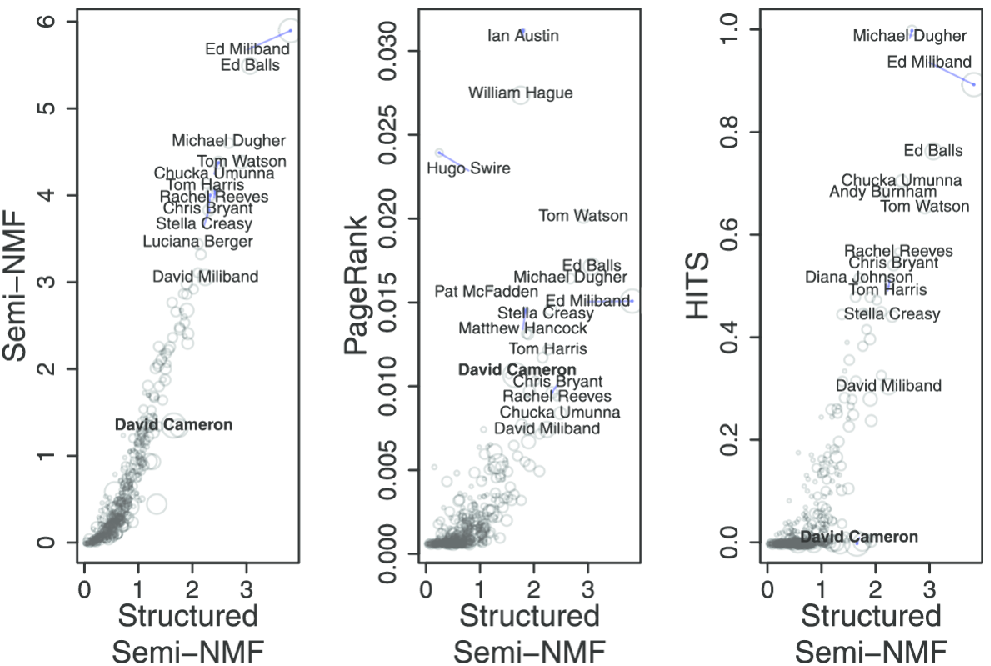}

\caption{Importance scores based on Structured Semi-NMF, Semi-NMF
($S_{m}=I_{n\times n}$), PageRank and HITS (Authority Scores).
PageRank and HITS are both calculated using the
Retweet network, while the other measures utilize all three networks.
The radius of the circle indicates the count of future newspaper
headlines as measured
with Lexis--Nexis. The top ten MPs for the methods in each
scatterplot are labeled. David Cameron, who is Prime Minister and in
boldface, was not in the top ten for any method.}
\label{fig:UKTwitter:rankings}
\end{figure}

Last, we discuss convergence criteria used for the ALS
algorithm. Let $\mathcal{O}^{(i)}$ denote the value of the objective
function at iteration $i$. Then the algorithm stops when
$\frac{|\mathcal{O}^{(i)} - \mathcal{O}^{(i-1)}|}{\mathcal{O}^{(i-1)}}
\le\varepsilon=10^{-4}$. We find in all our investigations that the
algorithm converges within $50$ iterations. $\varepsilon=10^{-4}$ is also
used for the projection threshold.

\section{Analysis of the political multiview Twitter
networks\texorpdfstring{.}{}} \label{sec:nmfResults}
\subsection{Does Twitter influence translate to the real
world\texorpdfstring{?}{?}}
Using the best rank six factorization with
$S_{m}$ defined with all four network statistics, we
rank MPs according to the estimated $\Theta$ and the importance
measure defined in equation~(\ref{eqn:importance}).

Figure~\ref{fig:UKTwitter:rankings} shows the importance scores from
the Structured Semi-NMF, Semi-NMF, PageRank and HITS. PageRank and
HITS are computed using the retweet network, which has been shown to
capture conversation dynamics better than the other network types
[\citet{cha2010measuring}]. Not surprisingly, the different importance
measures are all positively correlated.

Accordingly, as shown in Table~\ref{table:UKTwitter:rankings}, there
is general agreement between Structured Semi-NMF, Semi-NMF and HITS
in the top ten important MPs. Many of these MPs held leadership
positions in the coalition or Opposition cabinets. For instance,
\emph{Ed Miliband}, leader of the Labour Party and of the Opposition at
the time of writing,
is prominently emphasized in all rankings. \emph{Tom Watson} was the
Deputy Chair of the Labour Party, and \emph{Chuka Umunna} is the
Shadow Secretary of State for Business, Innovation and Skills. The
exceptions are \emph{Rachel Reeves}, who became the Shadow Secretary
of State
for Work and Pensions for the Opposition after the data was collected,
and \emph{David Miliband}, who held several important positions in previous
terms prior to data collection.

%
\begin{sidewaystable}
\tablewidth=\textwidth
\tabcolsep=0pt
\caption{MP rankings and in parentheses the party and frequency that
the MP appears in future headlines for Structured Semi-NMF, Semi-NMF
($S_{m}=I_{n\times n}$),
PageRank and HITS (Authority Scores). L~denotes Labour, C~denotes
Conservative} 
\label{table:UKTwitter:rankings}
%
\begin{tabular*}{\tablewidth}{@{\extracolsep{\fill}}@{}lcccc@{}}
\hline
\textbf{Rank} & \textbf{Structured Semi-NMF} & \textbf{Semi-NMF} &
\textbf{PageRank} & \textbf{HITS}\\
\hline
\phantom{0}1 & Ed Miliband (L, 2478) & Ed Miliband (L, 2478) & Ian Austin (L, 3) & Michael Dugher (L, 120)\\
\phantom{0}2 & Ed Balls (L, 580) & Ed Balls (L, 580) & William Hague (C, 771) & Ed Miliband (L, 2478)\\
\phantom{0}3 & Tom Watson (L, 253) & Michael Dugher (L, 120) & Hugo Swire (C, 57) & Ed Balls (L, 580)\\
\phantom{0}4 & Michael Dugher (L, 120) & Tom Watson (L, 253) & Tom Watson (L, 253) & Chuka Umunna (L, 203)\\
\phantom{0}5 & Chuka Umunna (L, 203) & Chuka Umunna (L, 203) & Ed Balls (L, 580) & Andy Burnham (L, 125)\\
\phantom{0}6 & Rachel Reeves (L, 54) & Rachel Reeves (L, 54) & Michael Dugher (L, 120) & Tom Watson (L, 253) \\
\phantom{0}7 & Stella Creasy (L, 178) & Chris Bryant (L, 164) & Pat McFadden (L, 1) & Rachel Reeves (L, 54)\\
\phantom{0}8 & Chris Bryant (L, 164) & Stella Creasy (L, 178) & Ed Miliband (L, 2478) & Chris Bryant (L, 164)\\
\phantom{0}9 & Tom Harris (L, 113) & Luciana Berger (L, 133) & Stella Ceasy (L, 178) & Diana Johnson (L, 105) \\
10 & David Miliband (L, 489) & Andy Burnham (L, 125)& Matthew Hancock (C, 32) & Tom Harris (L, 113)\\
\hline
\end{tabular*}
%
\end{sidewaystable}

Another commonality is that, with the exception of PageRank, every MP in
the top ten is from the Labour Party. Labour MPs tend to be estimated
as most important, followed by Conservative, and then Liberal Democrat
MPs. The relative ranking among parties is consistent with the data,
where Labour MPs tend to be the most active users in our data. Of the
top fifty Twitter accounts in terms of number of retweets or mentions,
only four are affiliated with another party---the Conservatives. The
Liberal Democrats are even less active, ranked in the hundreds in
terms of number of retweets or mentions. For instance, \emph{Nick
Clegg}, leader of the Liberal Democrats and Deputy Prime Minister at
the time of writing, is typically the top-ranked member of his party
at forty-nine with
Structured Semi-NMF, forty with PageRank, and outside the top hundred
with both Semi-NMF and HITS.

Activity in the data set is likely associated with longevity on
Twitter. For instance, \emph{David Cameron}, Prime Minister
and leader of the Conservatives, is ranked twenty-nine with Structured
Semi-NMF, sixty-eight with Semi-NMF, sixteen with PageRank, and
two hundred and forty-two with HITS. Cameron joined Twitter just as
the data was collected in October 2012, and, thus, may have
artificially low levels of activity when compared against more recent
data. In spite of these challenges, PageRank and Structured Semi-NMF
with use of the $S_{m}$ matrix are able to boost these key MPs
importance, even though they interact via Twitter with their MP
colleagues relatively infrequently.

We have so far seen anecdotal evidence that many MPs in leadership
positions are emphasized by the different techniques. Next, we test in
a regression setting whether these different measures of Twitter
importance predict media coverage, which is measured using Lexis--Nexis
(\surl{www.lexisnexis.com}) searches of the number of times an MP's name
appears in headlines from January~1, 2013, to October~17, 2013. This
interval of time is strictly after the Twitter data was collected to
avoid endogeneity issues. Because the headline counts were overdispersed,
we use a quasi-Poisson regression. The
mean and variance of the regression has form
%
\begin{eqnarray}
\label{eqn:headlineRegn} 
\mathbb{E}(\mbox{HeadlineCount}_{i}) &=& \exp(
\alpha+ \beta\mathcal{I}_{i} + \gamma\operatorname{Controls}_{i} ),
\\
\operatorname{Var}(\mbox{HeadlineCount}_{i}) &=& \rho\mathbb
{E}(\mbox{HeadlineCount}_{i}),
\label{eqn:headlineRegn2}
\end{eqnarray}
where $\rho\ge1$ is estimated from the data. HeadlineCount is the
headline occurrence frequency,
$\mathcal{I}$ is derived using the different importance measurement
techniques, and Controls contain the variables Age, Gender,
Constituency Size, Political Party and an indicator variable denoting
whether each MP represents a constituency within the city of London.
Age is an important control variable, since we expect younger MPs to
be more savvy with social media, which could affect their headline
coverage. Similarly, we expect MPs with larger constituencies,
certain political affiliations or London-based MPs to receive more
media attention.

Additional discussion in the supplemental article [\citet
{SupplementInfo}] shows the Poisson distributional
assumption appears more valid when compared to other distributions for
overdispersion, like negative binomial. Moreover, the
quasi-Poisson results featured the smallest root mean squared error
(RMSE) for all specifications that we discuss next.

\begin{figure}[t]

\includegraphics{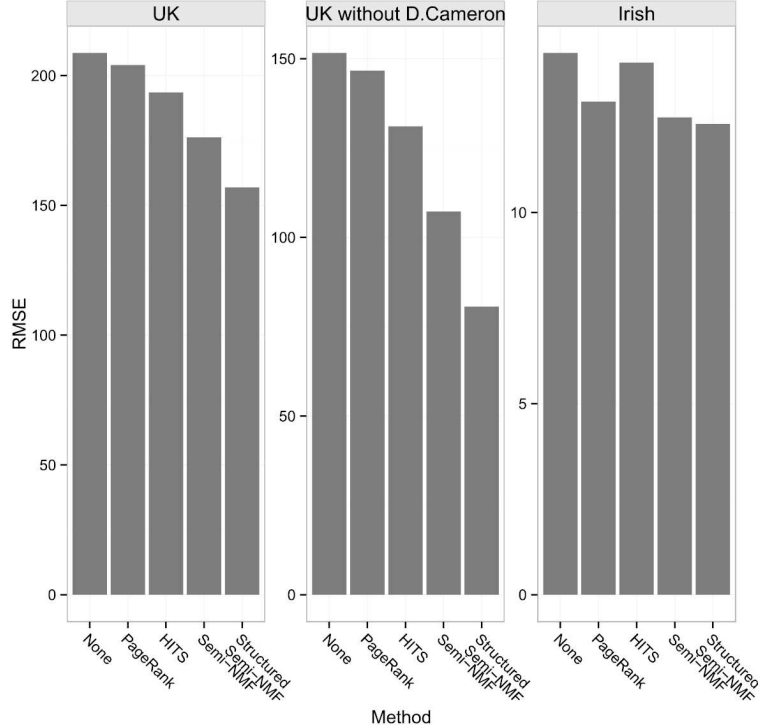}

\caption{Root mean squared errors for the predicted number of headlines
using different specifications of the
regression model in equations~(\protect\ref{eqn:headlineRegn}) and
(\protect\ref{eqn:headlineRegn2}). ``None'' refers
to including only control variables. ``PageRank'' refers to the control
variables plus the PageRank influence measure, ``HITS'' refers to
the control variables plus the HITS influence measure, and so on.}\label{fig:rmse}
\end{figure}

In Figure~\ref{fig:rmse}, we examine the RMSE of the model when using
only control variables, as
well as control variables with each influence measure separately. We
find that the model using the proposed factorization features the
lowest RMSE, especially after removing an outlier, David Cameron, who
received many more future headlines than predicted. As mentioned above,
David Cameron joined Twitter just as the original data set was collected,
potentially creating an artificially low presence on Twitter.

Table~1 in the supplemental article [\citet{SupplementInfo}]
shows the full
results for the estimated model with Structured Semi-NMF, where the
corresponding coefficient is statistically significant and positive as
expected. Specifying $S_{m}$ leads to an importance measure that is
associated with future media headlines even when controlling for
other influence measures and demographic information, thus
illustrating the importance of guiding the factorization solution.

%

\subsection{Identifying important conversation flows\texorpdfstring{.}{}}

%
\begin{figure}[t]
\begin{tabular}{@{}ccc@{}}

\includegraphics{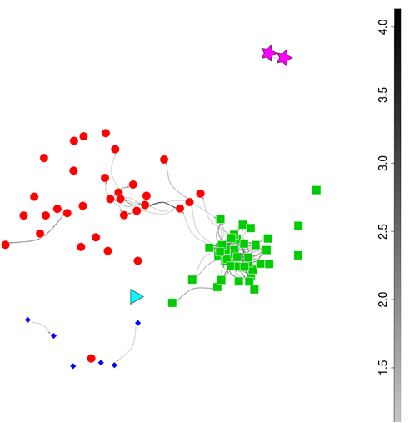}
 & \includegraphics{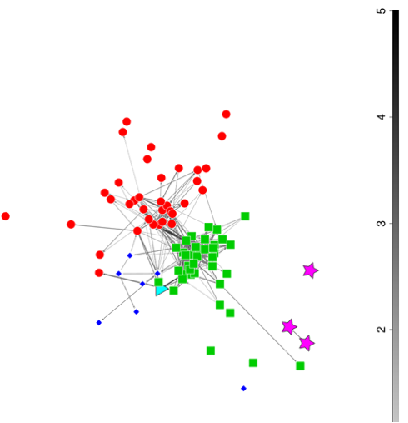} &\includegraphics{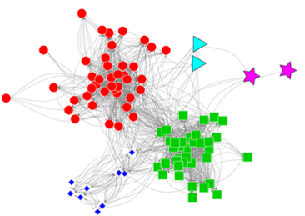}\\
\footnotesize{(a) Retweet network} & \footnotesize{(b) Mentions
network} & \footnotesize{(c) Follows network}
\end{tabular}
\caption{Subnetworks of UK Members of Parliament chosen by taking the
highest degree MPs in each party, with color and vertex shapes
denoting party affiliation. MPs are drawn in the same position as in
Figure~\protect\ref{fig:UKTwitter}.}
\label{fig:UKTwitter2}
\end{figure}

%
\begin{figure}[b]

\includegraphics{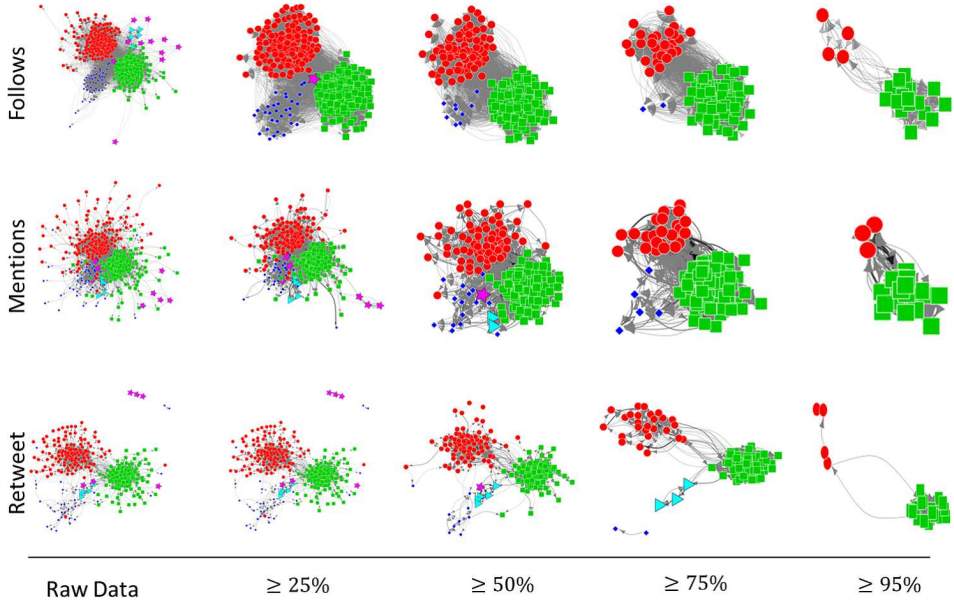}

\caption{Networks of UK Members of Parliament, with color and vertex
shapes denoting party affiliation. MPs in the top $q$th percentile of
$\sum_{k}(\Theta+ V_{m})_{ik}$ are kept and drawn in the
same position as in Figure~\protect\ref{fig:UKTwitter}.}
\label{fig:UKTwitter:snmf}
\end{figure}

Another advantage of the proposed factorization is that it can also be
used to extract potentially important conversation flows. We
construct subgraphs by keeping nodes in the top $q$th percentile of
$\sum_{k}(\Theta+V_{m})_{ik}$ to recover structure specific to each network
view.

The Structured Semi-NMF does not incorporate party affiliation for the
factorization. Yet it results in more interpretable subgraphs than
the alternative approach in Figure~\ref{fig:UKTwitter2} of looking at
high degree nodes within each party. Shown in
Figure~\ref{fig:UKTwitter:snmf}, there are denser within and between
party connections, and fewer isolated nodes. Moreover, with the
exception of a handful of MPs, each node can reach every other node on
the graphs. Thus, these networks help explain the influence rankings
from the previous section by identifying paths through which
interesting content flowed.

\begin{figure}
\begin{tabular}{@{}c@{\hspace*{9pt}}c@{\hspace*{9pt}}c@{}}

\includegraphics{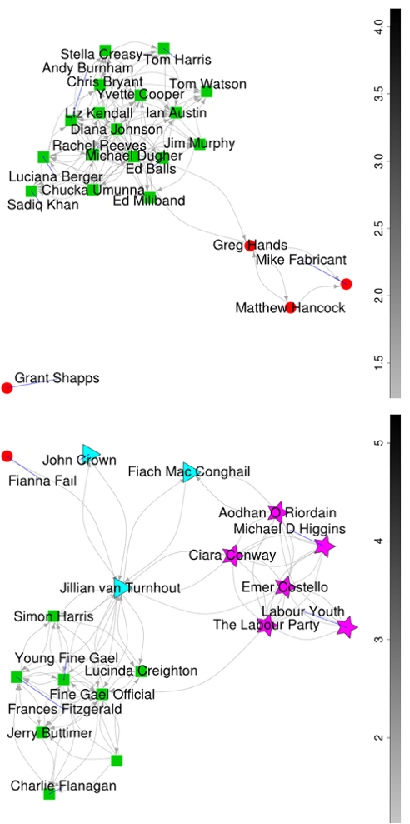}
 & \includegraphics{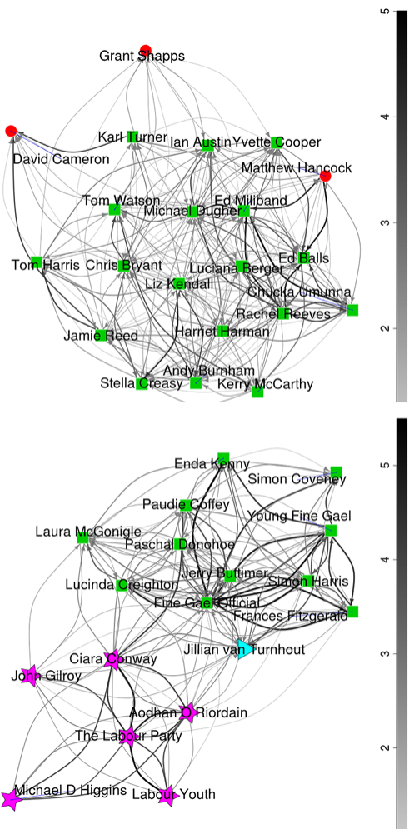} &\includegraphics{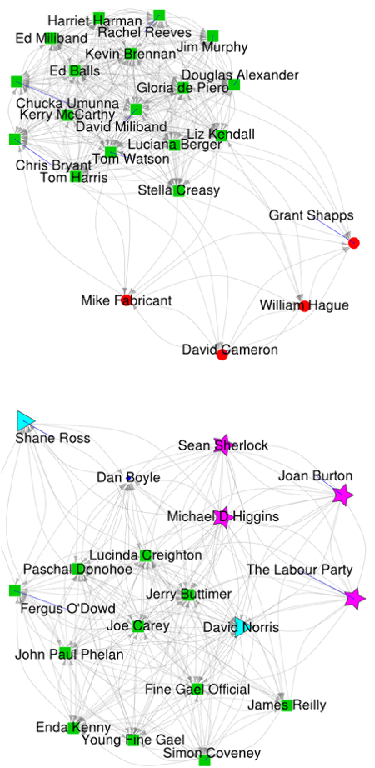}\\
\footnotesize{(a) Retweet network} & \footnotesize{(b) Mentions
network} & \footnotesize{(c) Follows network}
\end{tabular}
\caption{Subgraphs constructed for the UK MPs (top panel) and Irish
politicans (bottom panel), whose nodes are in the top $q=95$ percentile
of $\sum_{k}(\Theta+ V_{m})_{ik}$. Graphs are redrawn to optimize
vertex labels.}
\label{fig:snmf:layout}
\end{figure}

Tracing the flow of conversations in the $95$ percentile subgraphs in
Figure~\ref{fig:snmf:layout}, we see that the Labour politicians tend
to retweet each other often. Many of the Labour MPs, including
\emph{Stella Creasy}, \emph{Ed Miliband}, \emph{Chuka Umunna},
\emph{Rachel Reeves}, \emph{Tom Watson} and others, were universally
ranked as important in the previous section. \emph{Ed Balls} from
Labour interacts directly with \emph{Greg Hands} of the Conservative
party, who in turn forms a much smaller retweet clique with fellow
Conservatives \emph{Matthew Hancock} and \emph{Mike Fabricant}.

%

Since retweeting can amount to an endorsement, while mentioning allows
the author to control the content and sentiment, there are a greater
number of cross-party mentions edges. For instance, \emph{David
Cameron} is mentioned often and followed by Labour MPs, elevating
his importance on those specific networks, but is never
retweeted. This illustrates the value of utilizing all three types of
networks for measuring importance.

\subsection{Analysis of Twitter networks from the Irish political
sphere\texorpdfstring{.}{}} \label{sec:iedata}
We produce comparable, though less pronounced results with similar
Twitter network data from
the Irish political scene from late 2012. We organize the raw data
again provided in \citet{2013arXiv1301.5809G} into the same three
Twitter networks, each containing 348 nodes that represent the
accounts of Irish politicians and political organizations. The data
contains politicians from all levels of government, including the
President of the Republic of Ireland, members of the local and
national government, and elected representatives for the European
Union.

A majority of accounts belong to members of the Irish national
parliament, which is also a bicameral legislative body with elections
held at least once every five years using a system [\citet
{coakley2005politics}].
The lower house (D\'{a}il \'{E}ireann) is the principal
house in the Irish system and contains 166 elected members, the senate
(Seanad \'{E}ireann) contains a mixture of 60 appointed and elected
members. There are multiple political parties in the data: 33 Fianna
F\'{a}il, 127 Fine Gael, 6 Green, 20 Independent, 68 Labour, 22 Sinn
F\'{e}in and 8 Others. Approximately 60 Twitter accounts are
registered to political parties, for example, ``Fine Gael Official,''
``Labour Women,'' etc.

%
\begin{figure}[b]

\includegraphics{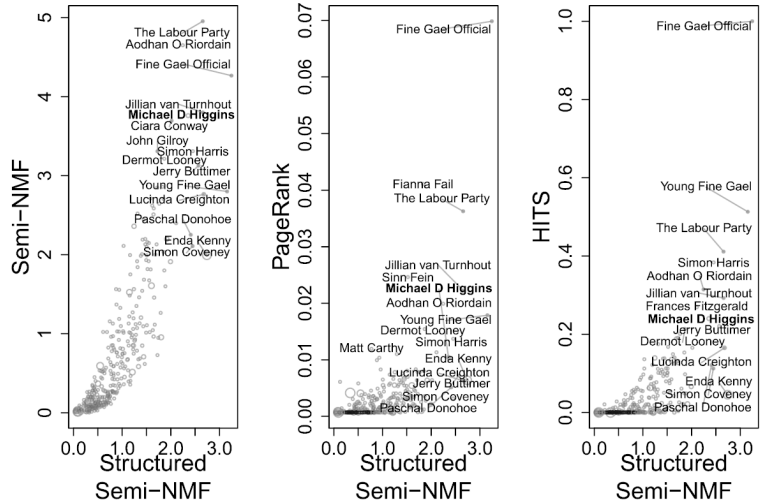}

\caption{Importance scores based on Structured Semi-NMF, Semi-NMF
($S_{m}=I_{n\times n}$), PageRank and HITS (Authority Scores) are both
calculated using the
Retweet network. The radius of the circle indicates count of future
newspaper headlines as measured
with Lexis--Nexis. The top ten Irish politicians for the methods in
each scatterplot are labeled. Michael Higgins, President, is
boldfaced.}
\label{fig:IETwitter:rankings}
\end{figure}

%
\begin{sidewaystable}
\tablewidth=\textwidth
\tabcolsep=0pt
\caption{Irish politician rankings and in parentheses the party and
frequency that the politician appears in future headlines for
Structured Semi-NMF, Semi-NMF ($S_{m}=I_{n\times n}$), PageRank and HITS
(Authority Scores). L denotes Labour, FG denotes Fine Gael, Ind
denotes Independent and SF denotes Sinn F\'{e}in. There are no
parenthetical headline counts or party names for political
organizations}\label{table:IETwitter:rankings}
%
\begin{tabular*}{\tablewidth}{@{\extracolsep{\fill}}@{}lcccc@{}}
\hline
\textbf{Rank} & \textbf{Structured Semi-NMF} & \textbf{Semi-NMF} &
\textbf{PageRank} & \textbf{HITS}\\
\hline
\phantom{0}1 & Fine Gael Official & The Labour Party & Fine Gael Official & Fine Gael Official\\
\phantom{0}2 & Young Fine Gael & Aodh\'{a}n \'{O} R\'{i}ord\'{a}in (L, 1) & Fianna F\'{a}il & Young Fine Gael\\
\phantom{0}3 & Enda Kenny (FG, 166) & Fine Gael Official & The Labour Party & The Labour Party \\
\phantom{0}4 & Lucinda Creighton (FG, 20) & Jillian van Turnhout (Ind, 0) & Sinn F\'{e}in & Simon Harris (FG, 4)\\
\phantom{0}5 & Jillian van Turnhout (Ind, 0) & Michael D Higgins (L, 25) & Jillian van Turnhout (Ind, 0) & Aodh\'{a}n \'{O} R\'{i}ord\'{a}in (L, 1)\\
\phantom{0}6 & The Labour Party & Ciara Conway (L, 0) & Aodh\'{a}n \'{O} R\'{i}ord\'{a}in (L, 1)& Jillian van Turnhout (Ind, 0)\\
\phantom{0}7 & Jerry Buttimer (FG, 2) & Simon Harris (FG, 4) & Young Fine Gael& Frances Fitzgerald (FG, 7) \\
\phantom{0}8 & Simon Harris (FG, 4) & John Gilroy (L, 3) & Dermot Looney (Ind, 0) & Michael D Higgins (L, 25) \\
\phantom{0}9 & Simon Coveney (FG, 10) & Dermot Looney (Ind, 0) & Simon Harris (FG, 4) & Jerry Buttimer (FG, 2) \\
10 & Paschal Donohoe (FG, 4) & Jerry Buttimer (FG, 2) & Matt Carthy (SF, 0)& Dermot Looney (Ind, 0)\\
\hline
\end{tabular*}
\end{sidewaystable}

After specifying $S_{m}$ as before and setting $K=7$ (chosen in a
similar fashion), we plot the importance scores in
Figure~\ref{fig:IETwitter:rankings} and list the top ten accounts in
Table~\ref{table:IETwitter:rankings} from the Structured Semi-NMF,
Semi-NMF, PageRank and HITS. In contrast to the British MP dynamics,
political organizations seem to play a much more important role in
online conversations within the Irish political sphere, as there is
broad agreement among the different importance measures that party
organization accounts are highly ranked, such as \emph{Fine Gael
Official}, \emph{Young Fine Gael}, and \emph{The Labour Party}. Some
politicians are also universally ranked as important. \emph{Michael D
Higgins}, the President at the time of writing, is ranked eleventh
under the Structured Semi-NMF, thirteenth under PageRank and in the
top ten for all other methods. \emph{Jillian van Turnhout} is an
appointed member of the Seanad \'{E}ireann and is consistently ranked
highly by the different influence measures. Likewise, \emph{Jerry
Buttimer} is a member of the D\'{a}il \'{E}ireann and formerly of
the Seanad \'{E}ireann, and \emph{Simon Harris} was elected to the
D\'{a}il \'{E}ireann in 2011 as its youngest member.

There are key differences, however, among the various importance
measures. \emph{Dermot Looney} is ranked in the top ten for Semi-NMF,
PageRank and HITS, but nineteenth under Structured Semi-NMF. He
seems to be ranked higher than one may expect, since Looney was part
of a local government and served as mayor of the South Dublin County
Council. \emph{Lucinda Creighton} is ranked fourth for the Structured
Semi-NMF, but is not in the top ten for other importance measures. At
the time of data collection, Creighton served as Minister for European
Affairs representing Ireland in negotiations on Ireland's EU/IMF
bailout and the hosting of Ireland's presidency of the European Union.
We also see that \emph{Enda Kenny}, an Irish Fine Gael politician who
has been the Taoiseach (prime minister) since March~2011, is ranked in
the top ten only under the Structured Semi-NMF approach. He is ranked
fortieth with Semi-NMF, thirty-fourth with PageRank and
seventy-second with HITS.

\begin{figure}

\includegraphics{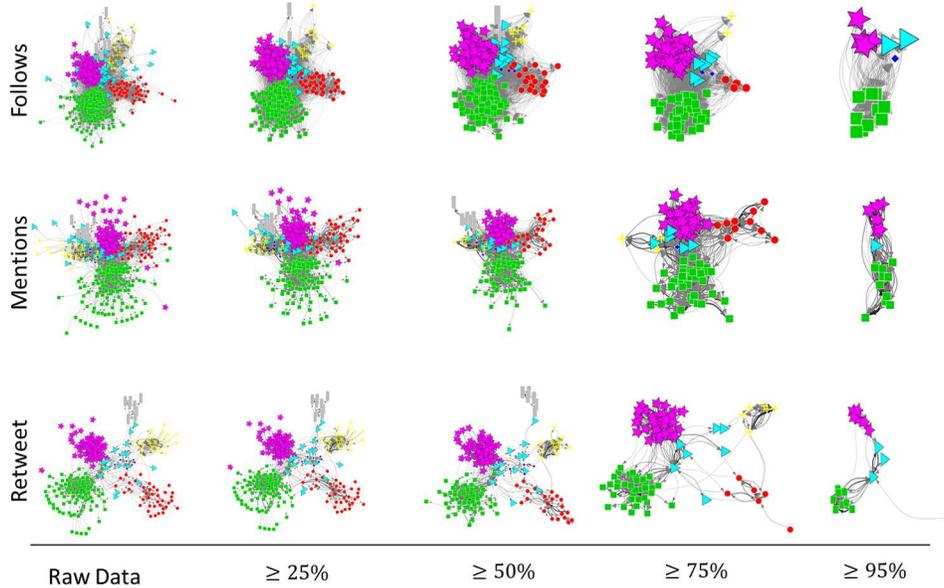}

\caption{Networks of Irish politicians, with color and vertex
shapes denoting party affiliation. Politicians in the top $q$th
percentile of
$\sum_{k}(\Theta+V_{m})_{ik}$ are kept and drawn in the
same position as in Figure~\protect\ref{fig:UKTwitter}.}
\label{fig:IETwitter:snmf}
\end{figure}

The larger differences between the Structured Semi-NMF and other
importance measures when compared to the UK MP results can be
explained by the sparser input networks, as shown in Figure~\ref
{fig:IETwitter:snmf}, which increase the effect of
the $S_{m}$ matrices. Figure~\ref{fig:snmf:layout} shows
the conversation dynamics that help explain why certain accounts are
ranked highly with the structured approach. For instance, we see that
\emph{Jillian van Turnhout}, an Independent, tends to be retweeted or
mentioned by Fianna F\'{a}il organizations in addition to Fine Gael,
Labour and other Independent politicians. Accounts within the Labour
party also form their own clique, centered around \emph{Michael D
Higgins} and the official Labour party account.

Finally, we test whether these different measures of Twitter
importance predict media coverage with the same quasi-Poisson model as in
equations~(\ref{eqn:headlineRegn}) and~(\ref{eqn:headlineRegn2}).
Headline occurrence frequency from
January 1, 2013, to October 17, 2013, is again measured using
Lexis--Nexis searches, $\mathcal{I}$ is derived using the different
importance measurement techniques, and Controls contains the variables
Age, Gender, Politician Type (local, presidential, D\'{a}il
\'{E}ireann, Seanad \'{E}ireann, European Union), Constituency and
Political Party. Since the data contains politicians in local
government, where, for example, exact constituency size is not easily
defined for council members, we include a fixed effect for every
unique electoral district or area. The 134 unique areas are
identified using a number of online sources, including official party
and candidate websites, newspaper articles and election results posted
on \surl{https://electionsireland.org/}. Party organization accounts
are removed when estimating the regression model.

%

Table~2 in the supplemental article [\citet{SupplementInfo}] shows
the Structured Semi-NMF
measure is again a statistically significant predictor for headline coverage
rate, after controlling for
all other variables, and Figure~\ref{fig:rmse} shows again that the
proposed approach results in an influence measure that improves
forecasting accuracy relative to alternative model specifications.

\section{Conclusion\texorpdfstring{.}{}} \label{sec:conclusion}
The Structured Semi-NMF performs best in both data sets, though the
improvement was only slight in the Irish context. The overall results
were driven by utilizing all three types of networks for measuring importance
and specifying the $S_{m}$ matrices to boost important politicians with
particular types of linkages.

One potential issue with the analysis is that Lexis--Nexis coverage of
non-US media and, in particular, the Irish media appears to be
imperfect. However, even with poor coverage, as long as it is
representative of the overall media landscape, then the reported
results will be meaningful. We are also unaware of other tools that can be
used for such searches. Another issue is that politicians may appear in
headlines
that reference their office, for example, ``the president.'' A more
comprehensive newspaper headline count is difficult to ascertain, but
could in future work provide further validation of the results
presented here.

Given that both data sets are exclusively link meta-data, our findings
support the notion that the significant challenges associated with
content analysis can often be complimented or avoided with network
analysis tools for tasks like identifying individuals influential
within social networking platforms. We believe this is partly
explained by the restriction of the population to politicians and
closely related organizations, which
ensures to some extent that the unobserved content is both homogeneous
and relevant.

A related problem of identifying emergence of key individuals,
communities or trends based on network data requires data collected
over time. Smoothing strategies, such as in \citet{PhysRevE.88.042812},
should be useful to extend the given model for network
time-series. We believe the proposed model can be useful for
applications in
marketing and e-commerce, where data is collected on ecosystems that
are close to a steady state. Otherwise, as we saw with David Cameron,
the model can mischaracterize the importance of key
individuals. Specific questions relating to path properties, such as
information diffusion [\citet{Twitterkleinberg}] or the spread of
epidemics [\citet{Twitterh1n1}], likely require additional
methods and
techniques specific to those subtopics.

There also has been recent work on a related problem when node features are
measured along with network data [\citeauthor{2013arXiv1306.4708F} (\citeyear{2013arXiv1306.4708F,hoffAOASmultiwayFA}), \citet{yang2013community}]. For
instance, one may
have access to demographic information or topics and themes of each
account's tweets as in \citet{2012arXiv1207.0017G}. While it appears
the proposed model could be useful in this setting, using external
covariates on the nodes to construct $S_{m}$ likely raises additional
issues that require care, such as variables being available for some,
but not all nodes. In this work, the node-level statistics are
``internally'' calculated directly from the network and, thus, will
always cover the full network.

A strength of the Structured Semi-NMF model is that it encompasses
different types of links (weighted and binary), integrates information
from multiple networks and allows the analyst
to utilize contextual knowledge about the given networked system. The
method depends upon the analyst choosing appropriate, context-specific
node-level statistics. As such, the alternating least squares
algorithm provides opportunities for additional regularization in
situations where the $S_{m}$ matrices are high dimensional or when
there are no node-specific values that are obvious to
use.



\begin{supplement}[id=suppA]
\stitle{Supplement to ``Analysis of multiview legislative networks
with structured matrix factorization: Does Twitter influence translate
to the real world?''}
\slink[doi]{10.1214/15-AOAS858SUPP} 
\sdatatype{.pdf}
\sfilename{aoas858\_supp.pdf}
\sdescription{We provide additional simulation results, details and
derivations for estimation algorithms, and detailed Poisson regression results.}
\end{supplement}

%
%

\printaddresses
\end{document}